\documentclass[showpacs,preprintnumbers,amsmath,amssymb,aps, prd]{revtex4}
\usepackage{graphicx}
\usepackage{grffile}
\usepackage{mathrsfs, mathtools}
\usepackage{caption}
\usepackage{float}
\usepackage{xcolor}
\usepackage{changepage}
\usepackage{dcolumn}
\usepackage{bm}
\usepackage{graphicx,subfigure,epsfig}
\usepackage{multirow}
\newcommand{\lagr}{\mathscr{L}}
\def\mn{_{\mu\nu}}
\def\MN{^{\mu\nu}}
\def\mN{_\mu^\nu}

\def\a{\alpha}
\def\b{\beta}
\def\f{\frac}
\def\rs{\rho_s}

\def\ad{\mathcal{A}_1(r)}
\def\ab{\mathcal{A}_2(r)}
\def\bd{\mathcal{B}_1(r)}
\def\bb{\mathcal{B}_2(r)}
\def\c{\cite}
\def\r{\ref}
\def\fr{f(r)}
\def\gr{g(r)}

\def\s{Schwarzschild }
\newcommand\be{\begin{equation}}
\newcommand\ee{\end{equation}}
\newcommand\ba{\begin{eqnarray}}
\newcommand\ea{\end{eqnarray}}
\newcommand\bt{\bibitem}
\newcommand\nn{\nonumber}
\newcommand\lt{\left}
\newcommand\rt{\right}
\newcommand\pt{\partial}
\newcommand\tx{\text}
\newcommand\mc{\mathcal}
\begin{document}
\title{Thermodynamics, Weak Gravitational Lensing, and Parameter Estimation of a \s Black Hole Immersed in Hernquist Dark Matter Halo}
\author{Sohan Kumar Jha}
\email{sohan00slg@gmail.com}
\affiliation{Department of Physics, Chandernagore College, Chandernagore, Hooghly, West
Bengal, India}

\date{\today}
\begin{abstract}
\begin{center}
Abstract
\end{center}
In this article, we obtain a novel black hole (BH) solution of a \s BH immersed in a Hernquist dark matter (SBHD)
halo. The thermodynamic properties of the resultant spacetime are then studied to gauge the impact of dark matter (DM) on the local and
global stability of the composite system of the BH-DM halo. With the intention of finding imprints of DM, we then studied weak gravitational lensing (GL) and
shadow. Both display significant dependence on the DM parameters - core radius $r_s$ and core density $\rs$. Finally, we constrain DM parameters
by utilizing bounds on the deviation parameter $\delta$ for super-massive BHs (SMBHs) $M87^*$ and $Sgr A^*$ reported by Event horizon telescope (EHT), Keck, and VLTI observatories. Our analysis
finds SBHD congruent with experimental observations, thereby making it a feasible candidate for an SMBH.
\end{abstract}
\maketitle
\section{Introduction}
The existence of dark matter is an intriguing topic that has fascinated researchers for quite some time now. Astrophysical BHs are considered to be surrounded by matter fields. One such plausible candidate for the matter field is DM. The first breakthrough in the quest for DM comes in observations of giant elliptical and spiral galaxies \cite{rubin}. According to one study by Persic et al., DM constitutes around $90\%$ of a galaxy's mass \c{persic}. Significant evidence points toward astrophysical BHs being embedded in DM halos \cite{akiyamal1,akiyamal6}. These revelations make it important to reckon DM's contribution near galactic center \c{sofue, boshkaye}. There exist different DM profiles that nicely incorporate the DM effect [\citenum{kiselev} - \citenum{rayimbaev}. Dehnen density profile, considered commonly for dwarf galaxies, encompasses different DM distributions depending on model parameters \c{dehnen, mo}. We will assess the effect of DM on different observations for the Dehnen-$(1,4,1)$ type, i.e., Hernquist DM distribution. Please see [\citenum{jusufi19} - \citenum{kalyan}] for recent works on the BH-DM halo composite system. Observations related to null geodesics play an important role in divulging important information related to the intrinsic nature of underlying spacetime. BH shadow draws the most attention in this regard. BH shadow is a dark $2d$ region in the celestial sky outlined by a bright ring. Due to its intrinsic dependence on BH parameters, it has been widely studied to extract valuable information and to seek evidence for the existence of DM [\citenum{RC}-\citenum{SN}].  \\
The thermodynamic properties of BHs provide a potent tool to probe their stability. The inception of Hawking radiation was a result of quantum mechanics being unified with general relativity (GR) \c{HAWKING}. These radiations are thermal fluctuations that are governed by pair productions that occur near the event horizon. One of the particles may escape the gravitational field of BH, constituting Hawking radiation [\citenum{HH}-\citenum{HC}]. These thermal fluctuations warranted the existence of a temperature associated with BH to be consistent with BH thermodynamics \c{BEK, KEIF}. Please see [\citenum{SW}-\citenum{SI}] for different techniques to obtain Hawking temperature. Information regarding the stability of BHs is embedded in specific heat and free energy. While the specific heat divulges information related to local stability, free energy encodes information with regard to global stability. Please see [\citenum{th1} - \citenum{th19}] for some of the recent notable works in this regard.\\
GL is an optical phenomenon related to null geodesics where the BH acting as a lens bends light rays from its path, leaving detectable imprints of BH properties on the deflection angle, thereby making it plausible to probe the existence of massive astronomical objects. GL also acts as a potent tool for studying the universe's expansion. Please see [\citenum{keeton}-\citenum{511}] for various studies related to GL. In this article, we mainly deal with weak GL with the help of Gauss-Bonnet theorem \c{GW, WERNER, ISHIHARA1, ISHIHARA2, ONO1, ONO2, ONO3}. Employing the method mentioned in \cite{CRISNEJO}, we derive the deflection angle with higher-order terms. The feasibility of theoretical models can be put to the test against experimental observations, and as such, observations related to various astrophysical phenomena equip us with an excellent tool. Observations concerning the shadow of SMBHs $M87^*$ and $Sgr A^*$ by EHT, Keck, and VLTI observatories present an unparalleled opportunity to test our model against experimental observations. In particular, bounds on the deviation parameter for $M87^*$ \c{M871, M872} and $Sgr A^*$ \c{keck, vlti1, vlti2} will be utilized to constrain free parameters. \\
We organize our article as follows. Section II is where we derive the line element for \s BH immersed in the Hernquist DM halo and discuss the nature of singularity and uniqueness of our model with the help of curvature invariants. Section III deals with thermodynamic properties, and section IV assesses the impact of DM on weak gravitational lensing. Section V studies shadow, and in section VI, we obtain constraints on the core radius and core density. We end this article with concluding remarks in section VII. We have used $G=c=M=1$ in this article.
\section{BH embodied in a Hernquist DM halo}
We start this section by considering the following DM profile \c{mo}
\begin{equation}
\rho(r) = \rho_s \left(\frac{r}{r_s}\right)^{-\gamma } \left[\left(\frac{r}{r_s}\right)^{\alpha }+1\right]^{\frac{\gamma -\beta }{\alpha }},
\label{dm}
\end{equation}
which, depending on the values of $(\a, \b, \gamma)$, models different DM distributions. $\gamma$ and $\b$ govern the dependence of the distribution at small and large $r$, whereas $\a$ determine the sharpness of the profile's transition. We consider the Hernquist distribution which corresponds to $(\a. \b, \gamma)\,=\,(1,4,1)$. The density for the Hernquist profile takes the following form:
\be
\rho(r)=\rs \lt(\f{r}{r_s}\rt)^{-1}\lt[1+\f{r}{r_s}\rt]^{-3}.\label{hd}
\ee
Here, $\rs$ and $r_s$ are the central density and the core radius of the DM halo, respectively. The mass profile for the DM distribution is
\begin{equation}
M_{H} = \int_0^r 4\pi \rho (r')r'^2 dr' = \frac{2 \pi  r^2 r_s^3 \rs}{ \left(r_s+r\right)^2}.
\label{mh}
\end{equation}
we then assume the spherically symmetric metric for a pure DM halo to be
\begin{equation}
ds^2 = -\mathcal{A}_1(r) dt^2 + \mathcal{B}_1(r)^{-1} dr^2 + r^2 (d\theta^2 + \sin^2 \theta d\phi^2),
\label{ldm}
\end{equation}
where $\mathcal{A}_1(r)$ is the redshift function and $\mathcal{B}_1(r)$ is the shape function. Utilizing the relation between the tangential velocity of a particle $v_t$ and the redshift function $\mathcal{A}_1(r)$ in one hand and the relation between $v_t$ and $M_H$ on the other, we can now obtain $\mathcal{A}_1(r)$ from the following equation \c{vt}:
\be
v_t^2 = \frac{M_H}{r}=  r \frac{d}{dr} \ln \sqrt{\mathcal{A}_1(r)}\quad \Rightarrow \mc{A}_1=e^ {\int \f{2M_H}{r^2}dr}.
\ee
Conjoining the above equation with Eq. (\r{mh}) and assuming $\mc{A}_1=\mc{B}_1$, we get, after retaining leading order terms, the following expression
\be
\mc{A}_1(r)=\mc{B}_1(r)=e^{-\f{4\pi \rs r_s^3}{r+r_s}} \approx 1-\f{4\pi \rs r_s^3}{r+r_s},
\label{ah}
\ee
From the Einstein field equations
\begin{equation}
R\mn - \frac{1}{2} R g\mn = \kappa^2 T\mn^H,
\label{ed}
\end{equation}
for the line element (\r{ldm}), we obtain the energy-momentum tensor $T\mn^H$ for the spacetime where $T\mN = g^{\nu \lambda} T_{\mu \lambda} = \text{diag} [-\rho, p_r, p, p]$. In Eq, (\r{ed}), $R\mn$ is the Ricci tensor, $R$ is the Ricci scalar, and $g\mn$ is the metric tensor for the line element (\r{ldm}). Utilizing the field equations (\r{ed}) provides the following equations:
\begin{equation}
\begin{aligned}
\kappa^2 T_t^{t (H)}= & \bd \left(\frac{1}{r} \frac{\mc{B}_{1}^{\prime}(r)}{\bd}+\frac{1}{r^2}\right)-\frac{1}{r^2}, \\
\kappa^2 T_r^{r (H)} &= \bd\left(\frac{1}{r^2} + \frac{1}{r} \frac{\mathcal{A}_{1}^{\prime}(r)}{\mathcal{A}_{1}(r)}\right) - \frac{1}{r^2}, \\
\kappa^2 T_\theta^{\theta (H)} &= \kappa^2 T_\phi^{\phi (H)} = \frac{1}{2} \bd \left( \frac{\mathcal{A}_{1}^{\prime \prime}(r) \ad - \mathcal{A}_{1}^{\prime 2}(r)}{\mathcal{A}_{1}^2(r)} + \frac{1}{2} \frac{\mathcal{A}_{1}^{\prime 2}(r)}{\mathcal{A}_{1}^2(r)} \right. \\& \left.\hspace{2cm}+ \frac{1}{r} \left( \frac{\mathcal{A}_{1}^{\prime}(r)}{\mathcal{A}_{1}(r)} + \frac{\mathcal{B}_{1}^{\prime}(r)}{\mathcal{B}_{1}(r)} \right) + \frac{\mathcal{A}_{1}^{\prime}(r) \mathcal{B}_{1}^{\prime}(r)}{2 \ad \bd} \right) .
\end{aligned}
\label{td}
\end{equation}
Next, we consider the following metric for the combined system of BH and DM halo:
\be
ds^2 = -\fr dt^2 + \gr^{-1} dr^2 + r^2 (d\theta^2 + \sin^2 \theta d\phi^2),\label{lmc}
\ee
where we write the functions as
\be
\fr=\ad+\ab\quad \text{and} \quad \gr=\bd+\bb.
\ee
We will obtain functions $\ab$ and $\bb$ from field equations (\r{ed}) conjoined with the following field equations for the combined system:
\begin{equation}
R\mn - \frac{1}{2} R g\mn = \kappa^2 \lt( T\mn^H+T\mn^{B}\rt),
\label{ec}
\end{equation}
where $T\mn^{B}$ is the energy-momentum tensor for the \s BH. Since there exists no non-zero component of $T\mn^{B}$, each component of the energy-momentum tensor for the metric (\r{ldm}) must be equal to the corresponding component for the line element (\r{lmc}). This yields the following equations whose solutions produce $\ab$ and $\bb$:
\begin{equation}
\begin{aligned}
& \left(\bd+ \bb\right)\left(\frac{1}{r^2}+\frac{1}{r} \frac{\mathcal{B}_{1}^{\prime}(r)+\mathcal{B}_2^{\prime}(r)}{\mathcal{B}_{1}(r)+ \mathcal{B}_2(r)}\right) = \mathcal{B}_{1}(r)\left(\frac{1}{r^2}+\frac{1}{r} \frac{\mathcal{B}_{1}^{\prime}(r)}{\mathcal{B}_{1}(r)}\right), \\
& \left(\mathcal{B}_{1}(r)+ \mathcal{B}_2(r)\right)\left(\frac{1}{r^2}+\frac{1}{r} \frac{\mathcal{A}_{1}^{\prime}(r)+\mathcal{A}_2^{\prime}(r)}{\mathcal{A}_{1}(r)+\mathcal{A}_2(r)}\right)= \mathcal{B}_{1}(r)\left(\frac{1}{r^2}+\frac{1}{r} \frac{\mathcal{A}_{1}^{\prime}(r)}{\mathcal{A}_{1}(r)}\right) .
\end{aligned}.
\label{field_eqs}
\end{equation}
Above equations with $\ad=\bd$ produces $\ab=\bb$. Since the metric (\r{lmc}) reverts back to that for a \s BH, we must have $\ab=\bb=-\f{2M}{r}$. With this, the metric for the combined system of BH and DM halo becomes
\be
ds^2 = -\fr dt^2 + \fr^{-1} dr^2 + r^2 (d\theta^2 + \sin^2 \theta d\phi^2),\label{final}
\ee
where $\fr=\gr=1-\f{2M}{r}-\f{4\pi \rs r_s^3}{r+r_s}$. The above metric has two singularities: one at $r=0$ and another at $\fr=0$. The solution of the equation $\fr=0$ provides the position of the event horizon as
\be
r_h=\frac{1}{2} \left(\sqrt{\left(-2 M-4 \pi  r_s^3 \rho _s+r_s\right){}^2+8 M r_s}+2 M+4 \pi  r_s^3 \rho _s-r_s\right),
\ee
which reduces to the \s case in the limit $r_s \rightarrow 0$. We explore the qualitative impact of the DM parameters $\rs$ and $r_s$ on the event horizon in Fig. (\r{event}). As evident from the figure, for a fixed value of $r_s$, $r_h \propto \rs$, whereas for a fixed value of the core density, we have $r_h \propto  r_s^3$.
\begin{figure}[H]
\begin{center}
\begin{tabular}{cc}
\includegraphics[width=0.4\columnwidth]{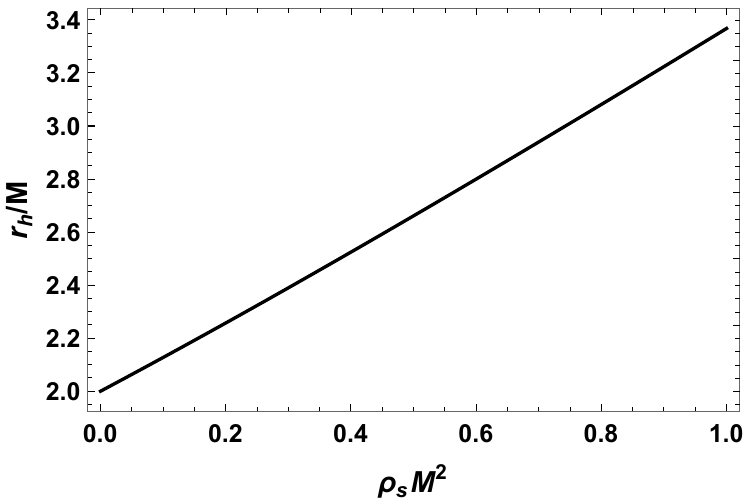}&
\includegraphics[width=0.4\columnwidth]{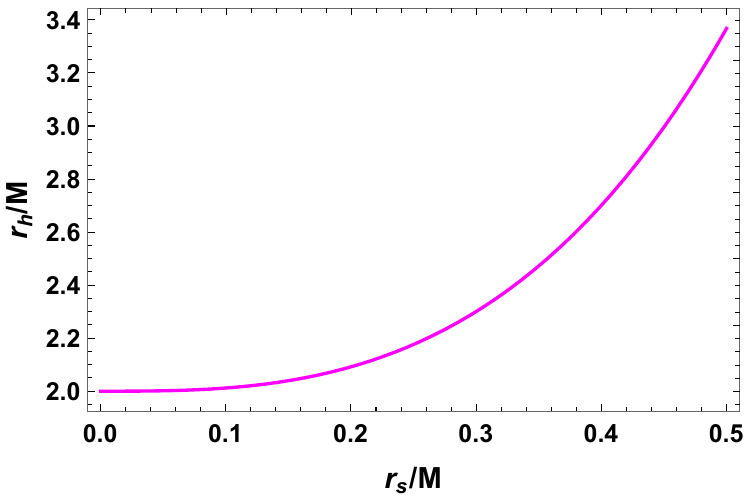}
\end{tabular}
\caption{Variation of the event horizon radius with the core density keeping $r_s=0.5M$ (left panel) and with the core radius keeping $\rs M^2=1.0$ (right panel). }\label{event}
\end{center}
\end{figure}
The curvature invariants will shed light on the nature of singularities at $r=0$ and $r=r_h$. Their expressions are given as follows:
\ba
&&\tx{Ricci Scalar}=R=\frac{8 \pi  r_s^5 \rho _s}{r^2 \left(r_s+r\right){}^3},\\\nn
&&\tx{Ricci squared}=R\mn R\MN=\frac{32 \pi ^2 r_s^8 \left(2 r^2+2 r r_s+r_s^2\right) \rho _s^2}{r^4 \left(r_s+r\right){}^6},\\\nn
&&\tx{Kretschmann Scalar}=K=\f{16}{r^6\lt(r+r_s\rt)^6}\\\nn
&&\lt(r^4 r_s^2 \left(45 M^2+76 \pi  M r_s^3 \rho _s+28 \pi ^2 r_s^6 \rho _s^2\right)+4 r^3 r_s^3 \left(15 M^2+15 \pi  M r_s^3 \rho _s+4 \pi ^2 r_s^6 \rho _s^2\right)+r^2 r_s^4
\left(45 M^2+24 \pi  M r_s^3 \rho _s+4 \pi ^2 r_s^6 \rho _s^2\right)\rt.\\\nn
&&\lt.+3 M^2 r_s^6+3 r^6 \left(M+2 \pi  r_s^3 \rho _s\right){}^2+6 r^5 r_s \left(M+2 \pi  r_s^3 \rho _s\right)
\left(3 M+2 \pi  r_s^3 \rho _s\right)+2 M r r_s^5 \left(9 M+2 \pi  r_s^3 \rho _s\right)\rt),
\ea
where the non-zero components of the Ricci tensor are
\ba
R_{tt}&=&\frac{4 \pi  r_s^4 \rho _s \left((r-2 M) r_s+r (r-2 M)-4 \pi  r r_s^3 \rho _s\right)}{r^2 \left(r_s+r\right){}^4},\\\nn
R_{rr}&=&\frac{4 \pi  r_s^4 \rho _s}{\left(r_s+r\right){}^2 \left(2 M r_s+2 M r-r^2+4 \pi  r r_s^3 \rho _s-r r_s\right)},\\\nn
R_{\theta \theta}&=&\frac{4 \pi  r_s^4 \rho _s}{\left(r_s+r\right){}^2},\\\nn
R_{\phi \phi}&=&\frac{4 \pi  \sin ^2(\theta ) r_s^4 \rho _s}{\left(r_s+r\right){}^2}.
\ea
It implies that the metric under consideration is not Ricci flat. While the curvature tensors diverge at $r=0$, no singularity exists at the event horizon. This makes the singularity at $r=r_h$ a coordinate singularity that can be circumvented by a suitable coordinate transformation. The singularity at $r=r_h$, however, is an essential singularity that can not be removed by a coordinate transformation. Even though the density profile of the DM halo is regular at $r=0$, the combined system of BH and DM halo is not. It is also to be noted that in the limit $r \rightarrow \infty$, above curvature invariants become zero. They clearly differ from the \s case. These facts signify the uniqueness of our model. Having explored the nature of singularities and the uniqueness of our model, we now move on to examining the thermodynamic stability of the BH.
\section{thermodynamic stability of \s bh in hernquist halo}
We gauge in this section the impact of DM on the thermodynamic stability of the combined system of \s BH and DM halo. To this end, various thermodynamic properties such as Hawking temperature, specific heat, and the Helmholtz free energy will be studied. We first express the Arnowitt-Deser-Misner (ADM) mass in terms of the event horizon utilizing the equation $f(r_h)=0$ as:
\be
M=\frac{r_h \left(r_h-4 \pi  r_s^3 \rho _s+r_s\right)}{2 \left(r_h+r_s\right)},\label{madm}
\ee
which in the limit $r_s \rightarrow 0$ reverts back to \s mass $M=\f{r_h}{2}$. It is to be noted that there exists a non-zero value of $r_h$, $r_h^M=4 \pi  r_s^3 \rho _s-r_s$ that makes the ADM mass zero. In the absence of DM halo, $r_h^M=0$. $r_h^M$ marks the lower limit of the event horizon. The Hawking temperature of the \s BH in Hernquist DM halo is
\be
T_{H}=\f{f'(r)}{4\pi}|_{r=r_h}=\frac{2 r_h r_s+r_h^2-4 \pi  r_s^4 \rho _s+r_s^2}{4 \pi  r_h \left(r_h+r_s\right){}^2},\label{th}
\ee
which in the limit $r_s \rightarrow 0$ reduces to $T_H=\f{1}{4\pi r_h}$, the temperature for a \s BH in vaccum. Evaporation due to Hawking radiation stops when $r_h$ reaches $r_h^M$. While the temperature at the end of the evaporation process takes the value of infinity in the case of a \s BH in vacuum, this temperature for a \s BH embodied in a Hernquist DM halo is finite, viz. $T_H^M=\frac{1}{16 \pi ^2 r_s^3 \rho _s}$. It would be instructive to explore the impact of DM parameters on temperature $T_H$. As such, In Fig. (\r{thab}), we display variation of $T_H$ with respect to $r_h$ for different values of $\rs$ and $r_s$.
\begin{figure}[H]
\begin{center}
\begin{tabular}{cc}
\includegraphics[width=0.4\columnwidth]{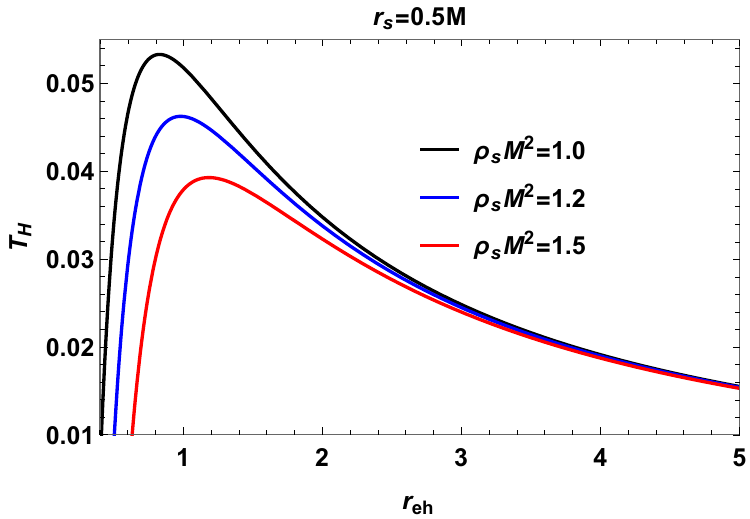}&
\includegraphics[width=0.4\columnwidth]{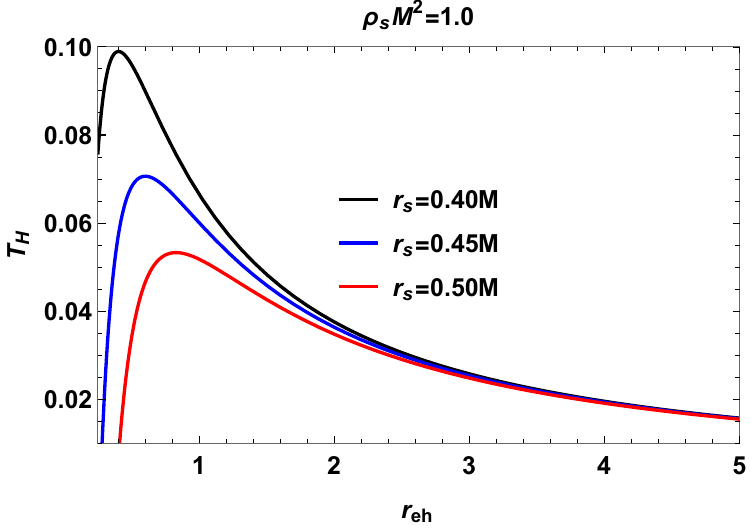}
\end{tabular}
\caption{Variation of the Hawking temperature with the core density kepping $r_s=0.5M$ (left panel) and with the core radius keeping $\rs M^2=1.0$ (right panel). }\label{thab}
\end{center}
\end{figure}
There are a few observations that one can readily make from Fig. (\r{thab}). First is the existence of a peak value $T_H^0$. Second is the diminishing value of $T_H^0$ with increasing either $\rs$ or $r_s$. Third is the position where the temperature peaks shifting towards a larger value of $r_h$ with increasing either $\rs$ or $r_s$. The peak signifies a local phase transition where the specific heat diverges. To shed more light on this, we calculate the specific heat for the system as follows:
\be
C_v=\f{\pt M}{\pt T_H}=-\frac{2 \pi  r_h^2 \left(r_h+r_s\right) \left(2 r_h r_s+r_h^2-4 \pi  r_s^4 \rho
_s+r_s^2\right)}{-12 \pi  r_h r_s^4 \rho _s+3 r_h r_s^2+3 r_h^2 r_s+r_h^3-4 \pi  r_s^5
\rho _s+r_s^3}.\label{cv}
\ee
The specific heat diverges at the critical radius
\be
r_h^c=\frac{2 \sqrt[3]{2} \pi ^{2/3} r_s^4 \rho _s}{\sqrt[3]{\sqrt{r_s^{10} \rho _s^2-4 \pi
r_s^{12} \rho _s^3}-r_s^5 \rho _s}}+2^{2/3} \sqrt[3]{\pi } \sqrt[3]{\sqrt{r_s^{10} \rho
_s^2-4 \pi  r_s^{12} \rho _s^3}-r_s^5 \rho _s}-r_s.
\ee
To accentuate the impact of DM parameters on specific heat, we display variation of specific heat with the event horizon for different values of core radius and density in Fig. (\r{cvab}). As evident from the figure, $C_v > 0$ for $r \in [r_h^M, r_h^c)$ signifying local stability of the BH-DM halo combined system, whereas $C_v < 0$ for $r_h > r_h^c$ forecasting thermodynamic instability. The critical radius increases with increasing either $\rs$ or $r_s$.
\begin{figure}[H]
\begin{center}
\begin{tabular}{cc}
\includegraphics[width=0.4\columnwidth]{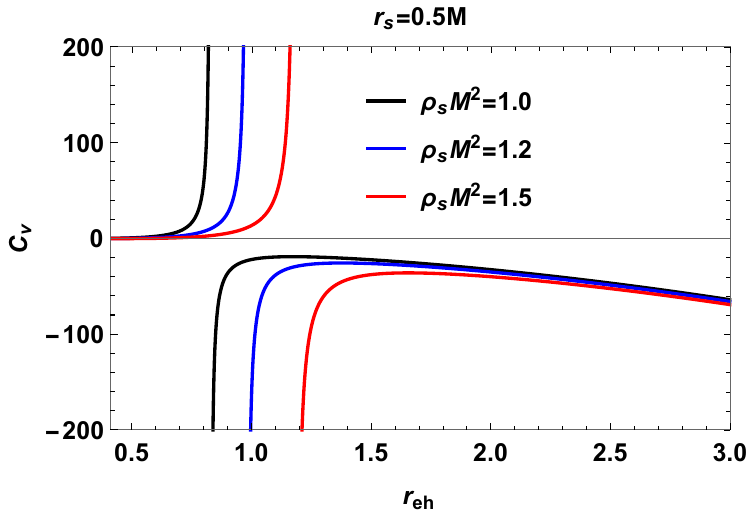}&
\includegraphics[width=0.4\columnwidth]{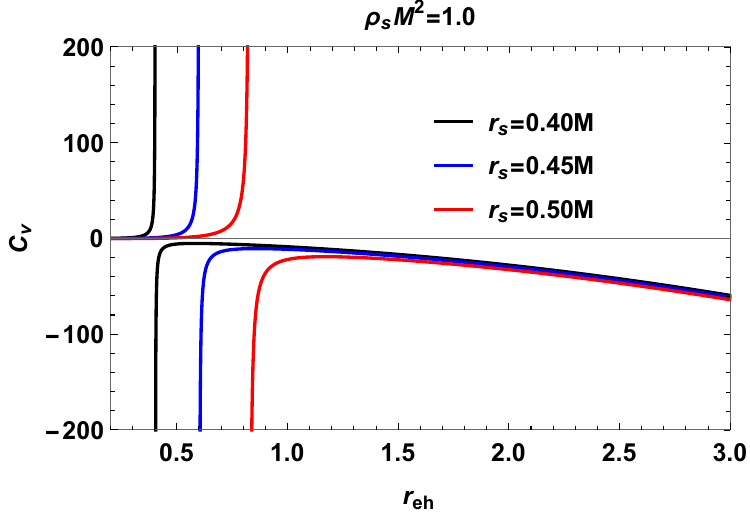}
\end{tabular}
\caption{Variation of specific heat with the core density kepping $r_s=0.5M$ (left panel) and with the core radius keeping $\rs M^2=1.0$. }\label{cvab}
\end{center}
\end{figure}
We move further and explore the global stability of the combined system with the help of Helmholtz free energy. To this end, we first calculate the entropy as:
\be
S=\int \f{dM}{T_H}=\pi r_h^2.\label{entropy}
\ee
Utilizing Eqs. (\r{madm}), (\r{th}), and (\r{entropy}), we obtain the expression of the free energy as
\be
F=\frac{r_h \left(-8 \pi  r_h r_s^3 \rho _s+2 r_h r_s+r_h^2-4 \pi  r_s^4 \rho
_s+r_s^2\right)}{4 \left(r_h+r_s\right){}^2}.
\ee
The position of the event horizon $r_h^f$ where the function becomes zero marks the point of global phase transition given by
\be
r_h^f=4 \pi  r_s^3 \rho _s+2 \sqrt{\pi } \sqrt{4 \pi  r_s^6 \rho _s^2-r_s^4 \rho _s}-r_s.
\ee
\begin{figure}[H]
\begin{center}
\begin{tabular}{cc}
\includegraphics[width=0.4\columnwidth]{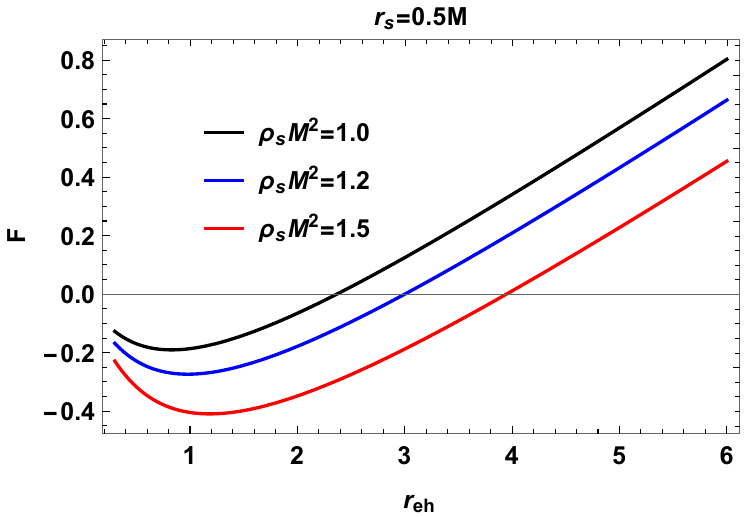}&
\includegraphics[width=0.4\columnwidth]{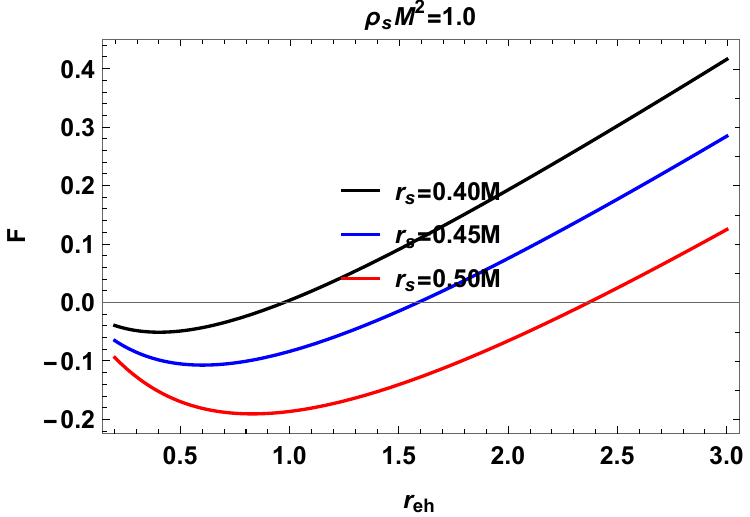}
\end{tabular}
\caption{Variation of free energy with the core density kepping $r_s=0.5M$ (left panel) and with the core radius keeping $\rs M^2=1.0$ (right panel). }\label{fab}
\end{center}
\end{figure}
As evident from Fig. (\r{fab}) free energy is negative for $r \in [r_h^M, r_h^f)$ indicative of global stability, whereas for $r> r_h^f$, the free energy is positive, implying global instability.We tabulate values of $r_h^M$, $r_h^c$, and $r_h^f$ below for different combinations of $(\rs M^2, r_s/M)$.
\begin{center}
\begin{tabular}{cccccc}
\hline
{}&$\rs M^2$ & $r_s/M$ & $r_h^{M}/M$ & $r_h^{c}/M$ & $r_h^{f}/M$\\
\hline\\
\multirow{4}{*}{}&{$0.4$}&{ }& $0.128319$ & $0.20479$ & $0.412264$\\[3mm]
&$0.8$&{} &$0.756637$ & $0.658757$ & $1.73174$\\[3mm]
&$1.2$&$0.5$&$1.38496$ & $0.980088$ & $3.00068$\\[3mm]
&$1.6$&{} & $2.01327$ & $1.24654$ & $4.2627$\\[3mm]
&$2.0$&{} & $2.64159$ & $1.47963$ & $5.52236$\\[3mm]
\hline\\
\multirow{4}{*}{}&{${}$}&$0.3$ & $0.039292$ & $0.0807252$ & $0.154754$\\[3mm]
&${}$&$0.4$ & $0.404248$ & $0.402827$ & $0.974437$\\[3mm]
&$1.0$&$0.5$ & $1.0708$ & $0.828541$ & $2.36772$\\[3mm]
&${}$&$0.6$ & $2.11434$ & $1.37374$ & $4.50996$\\[3mm]
&${}$&$0.7$ & $3.61027$ & $2.04047$ & $7.55503$\\[3mm]
\hline
\end{tabular}
\captionof{table}{Values of $r_h^M$, $r_h^c$, and $r_h^f$ for different values of DM parameters.}\label{critical}
\end{center}
We can infer from Table (\r{critical}) along with Figs. (\r{cvab}) and (\r{fab}) that a \s BH surrounded by a Hernquist halo of the higher core radius or density will have stability, both local and global, over a wider range of $r_h$ values.
\section{weak gravitational lensing}
\begin{figure}[H]
\begin{center}
\includegraphics[scale=0.5]{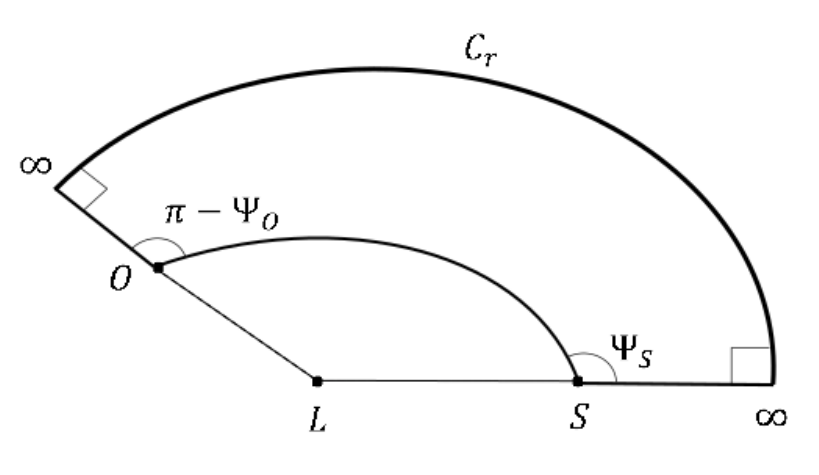}
\caption{Schematic diagram of the quadilateral ${}_O^{\infty}\Box_{S}^{\infty}$. }\label{lensing}
\end{center}
\end{figure}

Gravitational lensing is a potent tool that equips us to probe the intrinsic character of the background spacetime. Valuable information can be extracted from the deflection angle that bears the signature of the spacetime. This section is devoted to gauging the impact of the DM profile on the deflection angle, thereby exploring the feasibility of distinguishing a \s BH in vacuum from the one in DM halo. Equipped with the Gauss-Bonnet theorem \c{GW}, we will be examining the deflection angle using the formula \c{ISHIHARA1, CARMO}
\be
\gamma_D=-\int\int_{{}_R^{\infty}\Box_{S}^{\infty}} K
dS,\label{deflectionangle}
\ee
where ${}_O^{\infty}\Box_{S}^{\infty}$ is the quadrilateral displayed in Fig. (\ref{lensing}) and K is the Gaussian curvature given by
\ba\nn
K&=&-\f{\fr^{3/2}}{r}\f{d}{d\text{r}}\lt(\fr \f{d}{d\text{r}}(\f{r}{\sqrt{\fr}})\rt),\\\nn
&=&\frac{3 M^2}{r^4}-\frac{2 M}{r^3}+\left(\frac{12 \pi  M}{r^4}-\frac{4 \pi }{3 r^3}\right) r_s^3 \rho _s+\mc{O}\left(\f{\rs r_s^4}{r^4}, \f{M \rs r_s^4}{r^5} \rt).
\ea
Recasting Eq. (\ref{deflectionangle}) as \c{ONO1}
\be
\int\int_{{}_O^{\infty}\Box_{S}^{\infty}} K dS =
\int_{\phi_S}^{\phi_O}\int_{\infty}^{r_0} K \sqrt{\zeta}dr
d\phi,\label{Gaussian}
\ee
where $r_0$ is the least distance from BH and assuming a straight-line trajectory, we first calculate an initial value of the deflection angle, which comes out to be
\be
\gamma_D^0=\frac{3 \pi  M^2}{4 b^2}+\left(\frac{64 \pi  M^2}{9 b^3}+\frac{\pi ^2 M}{b^2}+\frac{8 \pi }{3 b}\right) r_s^3 \rho _s+\frac{4 M}{b}+\mc{O}\lt(\frac{M^3}{b^3},\frac{M^3\rs r_s^3}{b^4}\rt).
\ee
Higher order correction terms are obtained by utilizing the trajectory
\be
u=\f{1}{r}=\frac{\sin \phi }{b}+\frac{(1-\cos \phi )^2 \left(3 M+6 \pi  r_s^3 \rho
_s\right)}{3 b^2}-\frac{(3 \sin 3 \phi +60 \phi  \cos \phi-5 \sin \phi -32 \sin 2 \phi) \left(3 M+6 \pi  r_s^3 \rho _s\right){}^2}{144 b^3}+\mathcal{O}\left( \frac{M^3}{b^4}, \f{\rho_s r_s^4}{b^3}\right).\label{uorbit}
\ee
Using the above trajectory and integrating from $0$ to $\pi+\gamma_D^0$ in Eq. (\r{Gaussian}) yields the following expression for the deflection angle with higher order correction terms:
\be
\gamma_D=\frac{128 M^3}{3 b^3}+\frac{15 \pi  M^2}{4 b^2}+\left(\frac{243 \pi ^2 M^3}{b^4}+\frac{1648 \pi  M^2}{9 b^3}+\frac{9 \pi ^2 M}{b^2}+\frac{8 \pi }{3 b}\right) r_s^3 \rho
_s+\frac{4 M}{b}+\mc{O}\lt(\frac{M^4}{b^4},\frac{M^4\rs r_s^3}{b^5}\rt),\label{angle}
\ee
\begin{figure}[H]
\begin{center}
\begin{tabular}{cc}
\includegraphics[width=0.45\columnwidth]{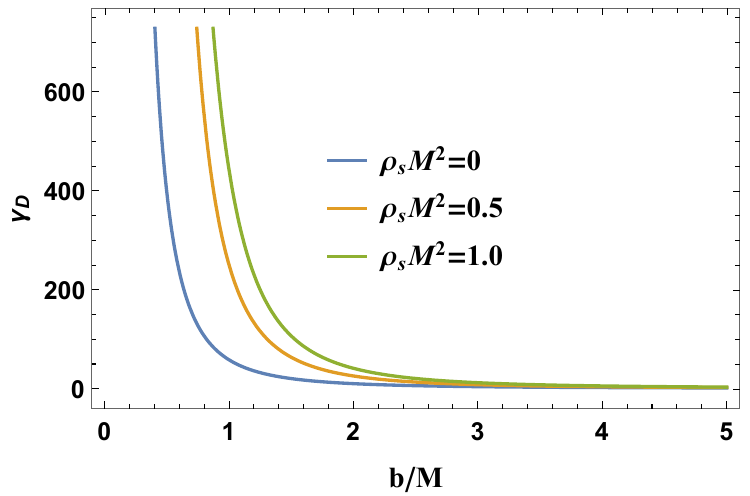}&
\includegraphics[width=0.45\columnwidth]{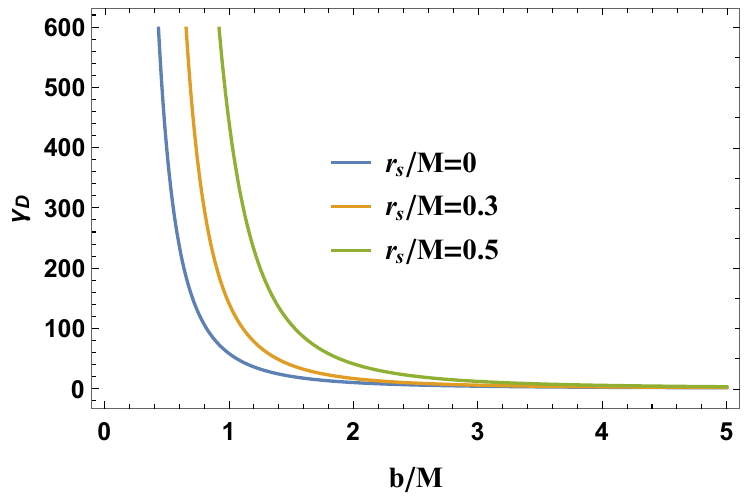}
\end{tabular}
\caption{Deflection angle for weak gravitational lensing. The left
the panel is for different values of core density keeping $r_s=0.5M$ and the right one is for different values of core radius keeping $\rs M^2=1.0$ .}\label{def}
\end{center}
\end{figure}
Fig. (\r{def}) demonstrates the impact of DM parameters on the deflection angle. It clearly shows that the deflection a light ray experiences while passing by a \s BH in Hernquist DM halo is significantly greater than while passing by a \s BH in a vacuum, thereby making it possible to differentiate the two on the basis of deflection angle. The accentuating effect of increasing either DM parameter on the deflection angle can also be inferred from Fig (\r{def}). We will next move on to the study shadow of SBHD.
\section{Shadow of a \s BH immersed in Hernquist DM halo}
This section is devoted to the study of shadows of SHDs that will act as a prologue for our final endeavor in this article, where we constrain DM parameters using shadow observables. As such, we will consider null geodesics on the equatorial plane $\theta=\f{\pi}{2}$. The first step in this regard is to consider the Lagrangian given by
\be
\lagr=\f{1}{2}(-\fr \dot{t}^2+\f{\dot{r}^2}{\fr}+r^2 \dot{\phi}^2),
\ee
where over-dot in the above equation is differentiation with respect to $\tau$, the affine parameter. Since the above metric does not feature any dependence on $t$ and $\phi$, their conjugate momenta $p_t$ and $p_\phi$, respectively, will be conserved, thereby making the energy $E$ and $L$ conserved quantities asl
\be
E=-p_t=-\f{\pt \lagr}{\pt \dot{t}}=\fr \dot{t} \quad \text{and} \quad L=p_\phi=\f{\pt \lagr}{\pt \dot{\phi}}=r^2 \dot{\phi}.
\ee
Utilizing $\dot{x_\mu}\dot{x^\mu}=0$ for mass-less particles conjoined with the above equations provides the following differential equation:
\be
\dot{r}^2=E^2 - \fr \lt(-\epsilon+\f{L^2}{r^2}\rt)=E^2 - V(r),
\ee
where the potential $V(r)$ governs the motion of the particle. Imposing conditions $\dot{r}=0$ and $\ddot{r}=0$ for circular orbits yield $V(r)=E^2$ and $\f{dV(r)}{dr}=0$, respectively. From the second equation, we get the following relation for the photo orbit of radius $r_{p}$:
\ba\nn
&&r_{p}f'(r_{p})=2f(r_{p}),\\\nn
&\Rightarrow&-6 M r_p r_s-3 M r_p^2-3 M r_s^2-4 \pi  r_p r_s^4 \rho _s-6 \pi  r_p^2 r_s^3 \rho _s+r_p r_s^2+2 r_p^2 r_s+r_p^3=0
\ea
The above equation has three roots: two imaginary and one real. The real root is the required photon radius given by
\ba
r_p&=&\frac{1}{3} \left(3 M+6 \pi  r_s^3 \rho _s-2 r_s\right)-\f{\sqrt[3]{2} }{3\sqrt[3]{P}}\left(3 \left(-6 M r_s-4 \pi  r_s^4 \rho _s+r_s^2\right)-\left(-3 M-6 \pi  r_s^3 \rho _s+2 r_s\right){}^2\right)+\f{\sqrt[3]{P}}{3\sqrt[3]{2}},
\ea
where
\ba\nn
P&=&54 M^3+324 \pi  M^2 r_s^3 \rho _s+54 M^2 r_s+648 \pi ^2 M r_s^6 \rho _s^2+18 M r_s^2+432 \pi ^3 r_s^9 \rho _s^3-216 \pi ^2 r_s^7 \rho _s^2+18 \pi  r_s^5 \rho _s+2 r_s^3\\\nn
&&+\lt(5832 \pi  M^3 r_s^5 \rho _s+19440 \pi ^2 M^2 r_s^8 \rho _s^2+5832 \pi  M^2 r_s^6 \rho _s+7776 \pi ^3 M r_s^{11} \rho
_s^3-10368 \pi ^2 M r_s^9 \rho _s^2+1944 \pi  M r_s^7 \rho _s\rt.\\\nn
&&\lt.-15552 \pi ^4 r_s^{14} \rho _s^4+11232 \pi ^3 r_s^{12} \rho _s^3-2700 \pi ^2 r_s^{10} \rho _s^2+216 \pi  r_s^8
\rho _s\rt)^{1/2}
\ea
The shadow of a non-rotating spherically symmetric BH is a circular region outlined by a bright ring whose radius $R_s$ is equal to the impact parameter associated with the photon orbit $b_p$ given by
\be
b_{ph}=R_s=\f{L}{E}=\f{r_{p}}{\sqrt{f(r_{p})}.}\label{Rs}
\ee
\begin{figure}[H]
\begin{center}
\begin{tabular}{cc}
\includegraphics[width=0.4\columnwidth]{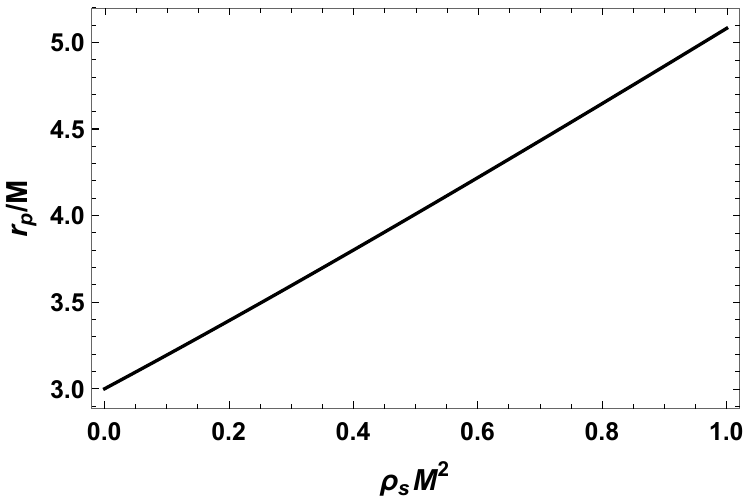}&
\includegraphics[width=0.4\columnwidth]{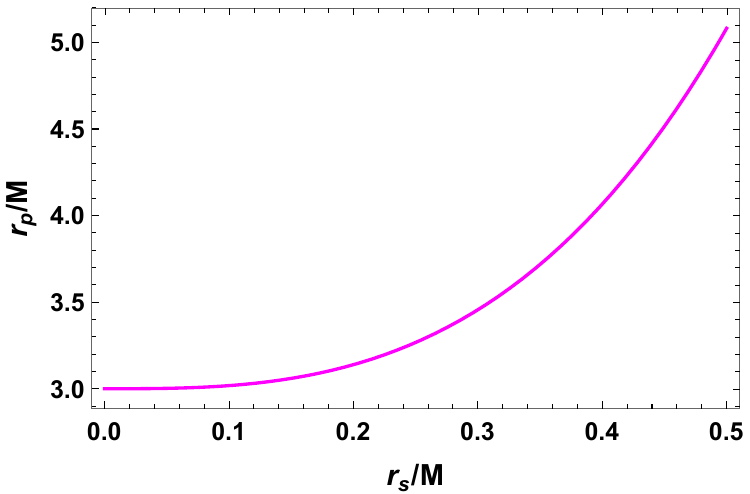}\\
\includegraphics[width=0.4\columnwidth]{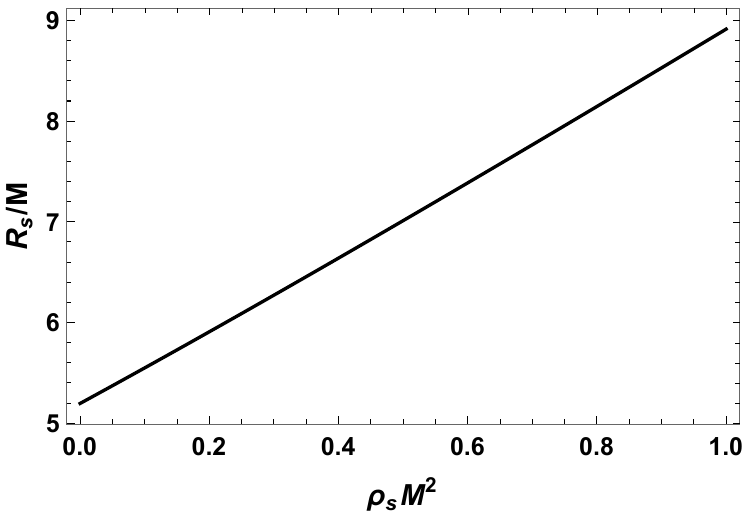}&
\includegraphics[width=0.4\columnwidth]{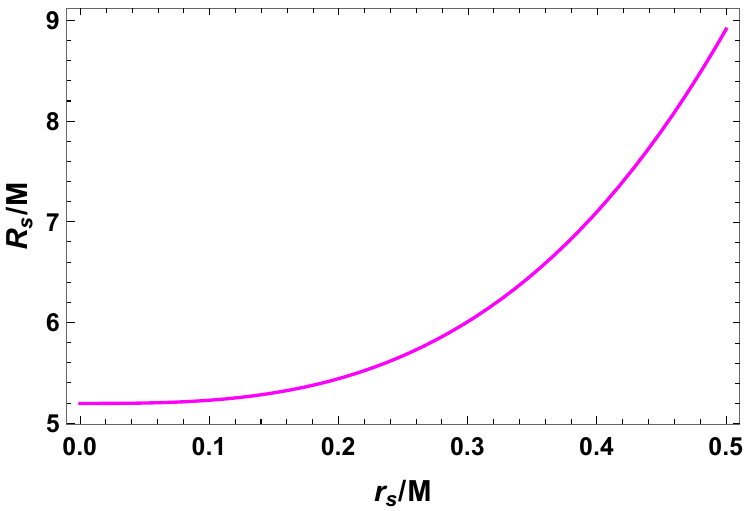}
\end{tabular}
\caption{Variation of photon radius $r_p$ and shadow radius $R_s$ with the core density kepping $r_s=0.5M$ (left panel) and with the core radius keeping $\rs M^2=1.0$ (right panel). }\label{rprs}
\end{center}
\end{figure}
We illustrate the dependence of photon radius and shadow radius on DM parameters in Fig. (\r{rprs}). Similar to the event horizon case, here, too, we see both $r_p$ and $R_s$ vary linearly with $\rs$ for a fixed value of $r_s$, whereas for a fixed value of the core density, we have $r_p\,(R_s) \propto  r_s^3$.
\begin{figure}[H]
\begin{center}
\begin{tabular}{cc}
\includegraphics[width=0.4\columnwidth]{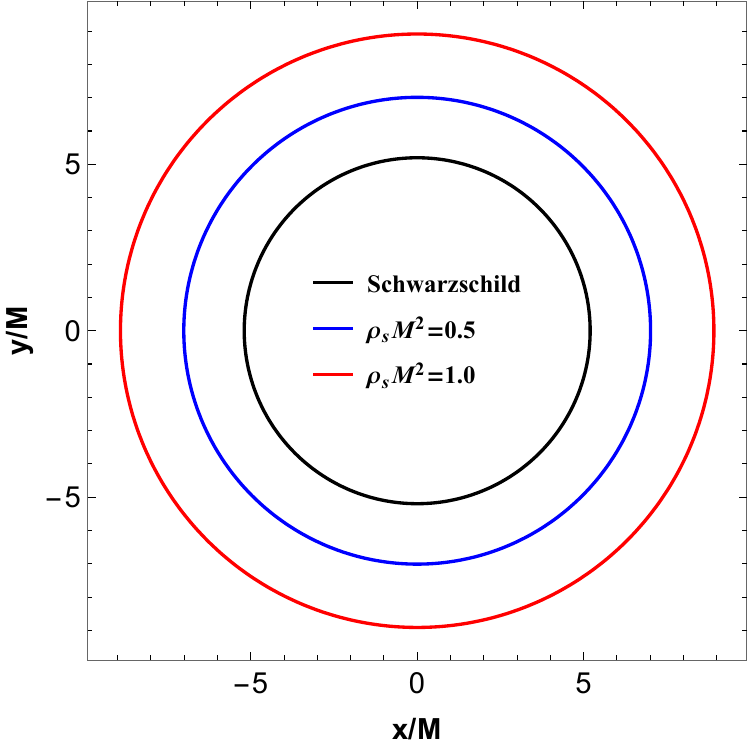}&
\includegraphics[width=0.4\columnwidth]{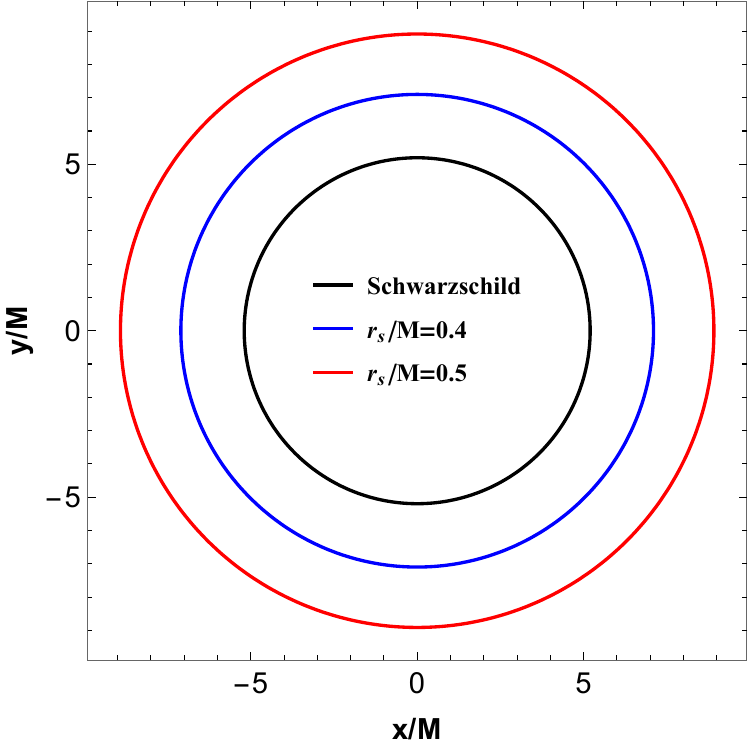}
\end{tabular}
\caption{BH shadows for different values of $\rs$ kepping $r_s=0.5M$ shown in left panel and for different values of $r_s$ kepping $\rs M^{2}=1.0$ shown in the right panel. }\label{shadow}
\end{center}
\end{figure}
Fig. (\r{shadow}) demonstrates shadows of a \s BH immersed in Hernquist halo for different values of core radius and density. The favorable impact of DM on the size of the shadow is conspicuous in the figure. Table (\r{rprsval}) provides a qualitative view of the change in the event horizon, photo radius, and shadow radius suffered due to a change in DM parameters.
\begin{center}
\begin{tabular}{cccccc}
\hline
{}&$\rs M^2$ & $r_s/M$ & $r_h/M$ & $r_p/M$ & $R_s/M$\\
\hline\\
\multirow{5}{*}{}&{$0.4$}&{ }& $2.52445$ & $3.80062$ & $6.63951$\\[3mm]
&$0.8$&{} &$3.08119$ & $4.64672$ & $8.14555$\\[3mm]
&$1.2$&$0.5$&$3.65831$ & $5.52086$ & $9.6892$\\[3mm]
&$1.6$&{} & $4.24864$ & $6.41292$ & $11.2565$\\[3mm]
&$2.0$&{} &  $4.84787$ & $7.31692$ & $12.8394$\\[3mm]
\hline\\
\multirow{5}{*}{}&{${}$}&$0.3$ &  $2.30015$ & $3.45582$ & $6.00737$\\[3mm]
&${}$&$0.4$ & $2.70049$ & $4.06552$ & $7.09918$\\[3mm]
&$1.0$&$0.5$ & $3.36773$ & $5.08105$ & $8.91373$\\[3mm]
&${}$&$0.6$ & $4.38782$ & $6.63036$ & $11.6673$\\[3mm]
&${}$&$0.7$ & $5.8496$ & $8.84477$ & $15.5798$\\[3mm]
\hline
\end{tabular}
\captionof{table}{Values of $r_h$, $r_p$, and $R_s$ for different values of DM parameters.}\label{rprsval}
\end{center}

\section{parameter estimation using shadow observable}
Experimental observations are potent tools to explore the feasibility of a proposed model and help constrain free parameters, which are core density $\rs$ and core radius $r_s$ in our case. To this end, we regard constraints on deviation from \s, $\delta$, reported by EHT for SMBH $M87^*$ and by Keck and VLTI observatories for SMBH $SgrA^*$ \cite{M871, M872, keck, vlti1, vlti2}. The observable $\delta$ is defined as \c{del}
\be
\delta=\f{R_s}{3\sqrt{3}M}-1,
\ee
where $R_s$ is the shadow radius for an SBHD defined in Eq. (\r{Rs}), $3\sqrt{3}M$ being the shadow radius in the absence of DM halo. Since the shadow radius for an SBHD is always greater than a \s BH in vacuum, we will always get a positive value of $\delta$ for the non-zero value of DM parameters. Reported bounds on $\delta$ for SMBHs $M87^*$ and $Sgr A^*$ are tabulated in (\ref{bounds}).
\begin{center}
\begin{tabular}{|l|c|c|c|r|}
\hline
BH & Observatory & $\delta$ & 1$\sigma$ bounds & 2$\sigma$ bounds\\
\hline
\hline
$M87^*$ & EHT & $-0.01^{+0.17}_{-0.17}$ & $4.26\le \frac{R_s}{M}\le 6.03$ &  $3.38\le \frac{R_s}{M}\le 6.91$\\
\hline
\hline
\multirow{2}{*}{$Sgr A^*$}&{VLTI} & $-0.08^{+0.09}_{-0.09}$ &{$4.31\le \frac{R_s}{M}\le 5.25$} &{$3.85\le \frac{R_s}{M}\le 5.72$}\\[3mm]
& Keck & $-0.04^{+0.09}_{-0.10}$ & {$4.47\le \frac{R_s}{M}\le 5.46$} & {$3.95\le \frac{R_s}{M}\le 5.92$}\\[1mm]
\hline
\end{tabular}
\captionof{table}{Bounds on $\delta$ from different observatories.} \label{bounds}
\end{center}
Since for SBHDs $\delta>0$, we will only consider upper bounds on $\delta$ reported by EHT, Keck, and VLTI to constrain $\rs$ and $r_s$. Fig. (\r{del1}) shows the variation of deviation parameter $\delta$ with core density, whereas its variation against core radius is illustrated in Fig. (\r{del2}). Horizontal lines in Fig. (\r{del}) are the upper bounds on $\delta$ as reported by various observatories. Interestingly, similar to the event horizon case, the deviation parameter too increases linearly with $\rs$, and its dependence on $r_s$ is of the cubic order.
\begin{figure}[H]
\centering \subfigure[]{ \label{del1}
\includegraphics[width=0.45\columnwidth]{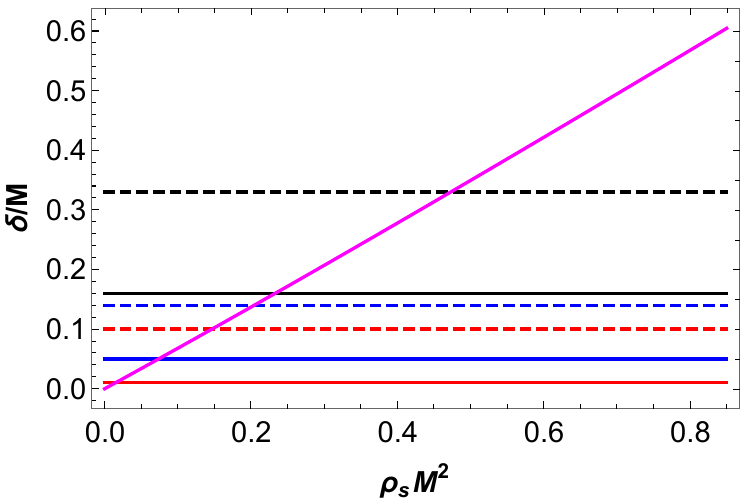}
} \subfigure[]{ \label{del2}
\includegraphics[width=0.45\columnwidth]{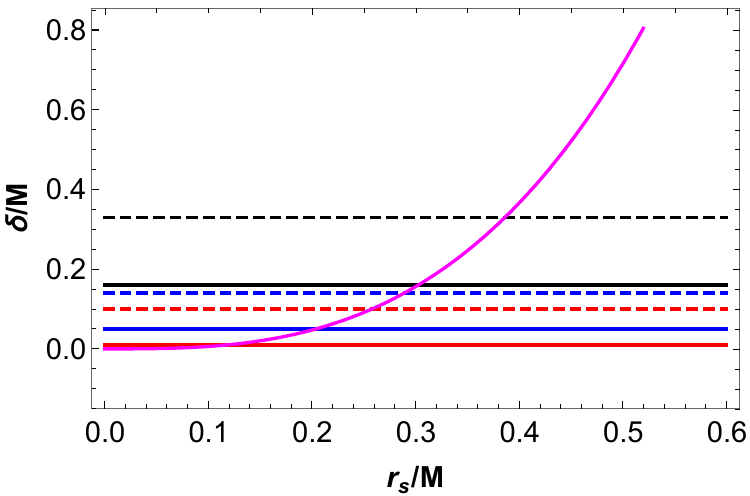}
}
\caption{Variations of deviation parameter shown with respect to core density keeping $r_s=0.5M$ in the left panel and with respect to core radius keeping $\rs M^2=1.0$ in the right panel. Horizontal solid lines are 1-$\sigma$ upper bound, and dashed lines are 2-$\sigma$ upper bounds. The color code for horizontal lines is: Black is for EHT, Blue is for Keck, and Red is for VLTI. }\label{del}
\end{figure}
The point where the $\delta$ variation line intersects the horizontal line marks the upper bound of the respective DM parameter. Our analysis shows that for $\rs M^2=1.0$, $r_s \in [0, 0.302507M]$ and $r_s \in [0, 0.38625M]$ within $1 \sigma$ and $2 \sigma$ confidence level (CL), respectively, for $M87^*$. Bounds from Keck observatory yield $r_s \in [0, 0.203867M]$ within $1 \sigma$ and $r_s \in [0, 0.289123M]$ within $2 \sigma$, whereas VLTI observations produce $r_s \in [0, 0.118254M]$  within $1 \sigma$ and $r_s \in [0, 0.257937M]$ within $2 \sigma$. We move further and obtain upper bounds on $\rs$ for $r_s=0.5M$. From observation related to $M87^*$ we obtain $\rs M^2 \in [0, 0.233074]$ ($1 \sigma$ CL) and $\rs M^2 \in [0, 0.473069]$ ($2 \sigma$ CL). For Keck we get $\rs M^2 \in [0, 0.0737781]$ ($1 \sigma$ CL) and $\rs M^2 \in [0, 0.204383]$ ($2 \sigma$ CL), whereas VLTI observation forecasts $\rs M^2 \in [0, 0.0148339]$ ($1 \sigma$ CL) and $\rs M^2 \in [0, 0.146654]$ ($2 \sigma$ CL). Obtained bounds from our analysis are tabulated in [\r{bounds1}].
\begin{center}
\begin{tabular}{ccccccccc}
\hline
{}&\multicolumn{3}{c}{} &\multicolumn{2}{c}{bounds on $r_s/M$} &{} &\multicolumn{2}{c}{bounds on $\rs M^2$}\\
\hline\\
{}&$\rs M^2$ &BH & Observatory &  1$\sigma$ bounds & 2$\sigma$ bounds &$r_s/M$  &1$\sigma$ bounds & 2$\sigma$ bounds\\
\\
\multirow{3}{*}{}&{}&{$M87^*$}& EHT& $0.302507$ & $0.38625$ &{} &$0.233074$&$0.473069$\\[3mm]
&$1.0$ &${}$&{Keck} &$0.203867$ & $0.289123$ &$0.50$ &$0.0737781$&$0.204383$\\[3mm]
&{}& $SgrA^*$&${}$&${}$ & ${}$ & ${}$&{}&{}\\[3mm]
&{} &${}$&{VLTI} & $0.118254$ & $0.257937$ &{} &$0.0148339$&$0.146654$\\[3mm]
\hline\\
\multirow{3}{*}{}&{}&{$M87^*$}& EHT& $0.284208$ & $0.362821$ &{} &$0.443731$&$0.903228$\\[3mm]
&$1.2$ &${}$&{Keck} &$0.191612$ & $0.271646$ &$0.40$ &$0.140143$&$0.388959$\\[3mm]
&{}& $SgrA^*$&${}$&${}$ & ${}$ & ${}$&{}&{}\\[3mm]
&{} &${}$&{VLTI} & $0.111198$ & $0.242373$ &{} &$0.0281515$&$0.278872$\\[3mm]
\hline
\end{tabular}
\captionof{table}{Bounds on $r_s/M$ and $\rs M^2$ from different observatories.} \label{bounds1}
\end{center}
Variation of deviation parameter with $\rs$ and $r_s$ for $M87^*$ and $Sgr A^*$ are elucidated in Figs. (\r{para1}) and (\r{para2}).
\begin{figure}[H]
\begin{center}
\begin{tabular}{cccc}
\includegraphics[width=0.4\columnwidth]{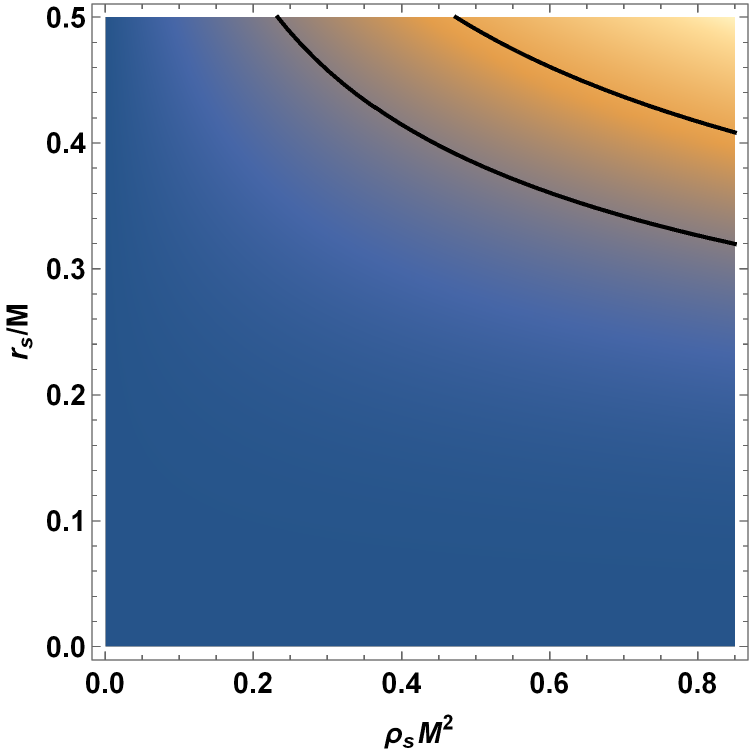}&
\includegraphics[width=0.05\columnwidth]{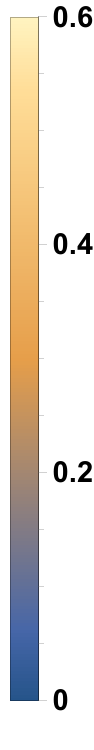}&
\includegraphics[width=0.4\columnwidth]{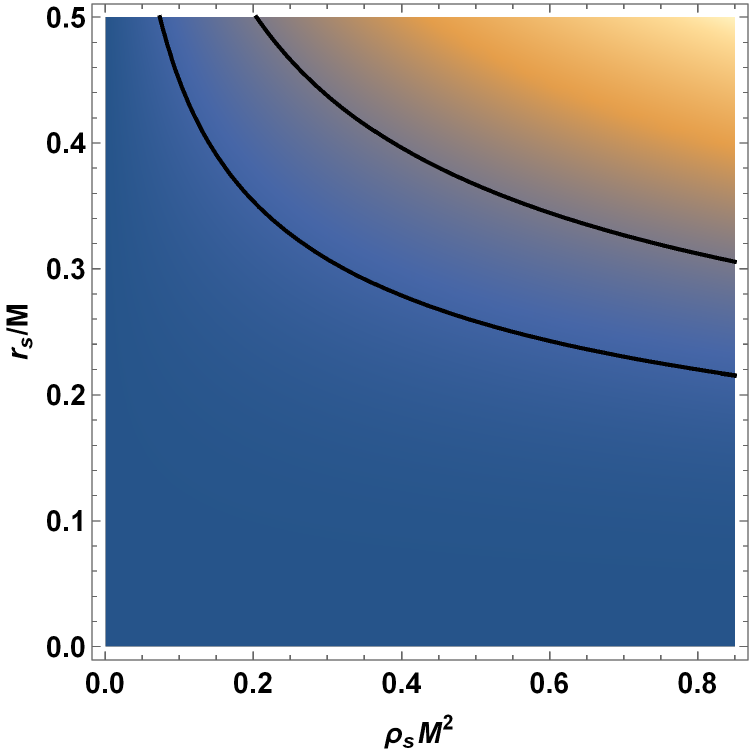}&
\includegraphics[width=0.05\columnwidth]{m87_bar.pdf}
\end{tabular}
\caption{Variation of deviation parameter $\delta$ with core $\rs$ and $r_s$. The left one is for $M87^*$ with EHT constraints, and the right one is for $Sgr A^*$ with Keck bounds. In each plot, the upper solid black line corresponds to the upper $2\sigma$ bound, and the lower one is for the upper $1 \sigma$ bound. }\label{para1}
\end{center}
\end{figure}

\begin{figure}[H]
\begin{center}
\begin{tabular}{cc}
\includegraphics[width=0.4\columnwidth]{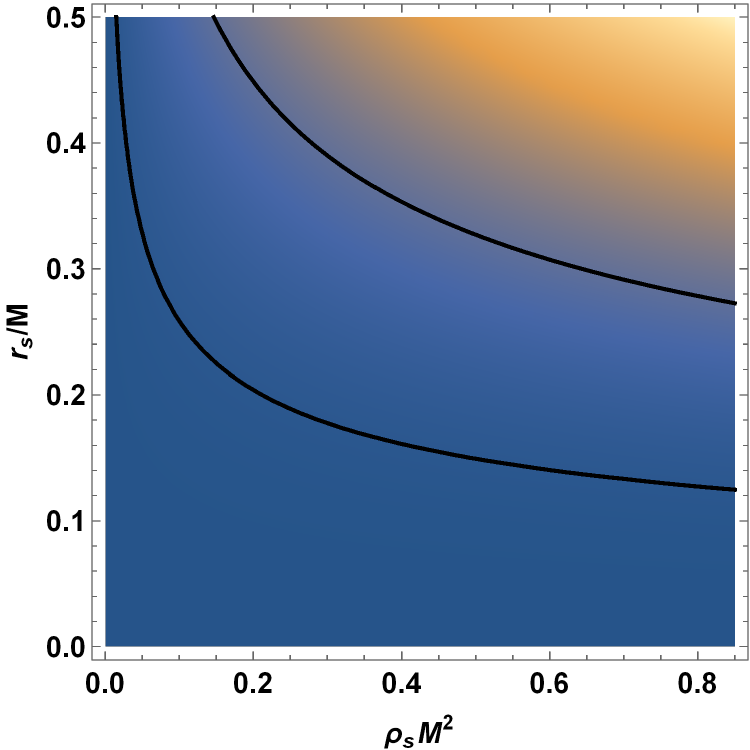}&
\includegraphics[width=0.05\columnwidth]{m87_bar.pdf}
\end{tabular}
\caption{Variation of deviation parameter $\delta$ with core $\rs$ and $r_s$ for $Sgr A^*$ with VLTI bounds. The upper solid black line corresponds to the upper $2\sigma$ bound, and the lower one is for the upper $1 \sigma$ bound. }\label{para2}
\end{center}
\end{figure}
Figs. (\r{para1}) and (\r{para2}) clearly exhibit congruence of our model with experimental observations for finite parameter space $(\rs M^2 - r_s/M)$, thereby making it a feasible candidate for an SMBH.
\section{conclusions}
This article studies the thermodynamic and optical properties of a \s BH embedded in Hernquist DM halo. We first obtain the line element for SBHD and its curvature invariants. The singularity of curvature invariants at $r=0$ is conspicuous, whereas no singularity exists where the lapse function becomes zero, i.e., at the event horizon. Hence, similar to the \s BH in vacuum, a \s BH immersed in Hernquist DM halo is not regular at $r=0$. Moreover, the metric for SBHD is not Ricci flat since non-zero components of Ricci tensors exist. We have also obtained expression for the event horizon and displayed its variation with respect to core radius and core density in Fig. (\r{event}). The dependence of the event horizon on the core density is found to be linear, whereas its reliance on the core radius is of cubic order. \\
Next, we explore the thermodynamic properties of SBHD. We first examined the Hawking temperature and displayed its variation with $r_h$ for different values of $\rs$ and $r_s$ in Fig. (\r{thab}). Significantly, our study revealed the occurrence of a finite temperature $T_H^M=\frac{1}{16 \pi ^2 r_s^3 \rho _s}$ at the end of the evaporation process when the ADM mass reduces to zero and the event horizon takes the value $r_h^M$. We readily observed from the figure the diminishing value of the peak and shifting of the peak position towards higher values of $r_h$ for higher values of either $\rs$ or $r_s$. The significance of the peak is that it marks a local phase transition where the specific heat diverges. Our next stop naturally becomes the specific heat. Specific heat divulges valuable information regarding the thermal stability of a system. Our analysis revealed that for $r_h \in [r_h^M, r_h^c]$, where $r_h^c$ is the position of the divergence of the specific heat, specific heat is positive, implying a thermally stable BH-DM halo system, whereas for $r_h > r_h^c$ we have negative values of specific heat, forecasting a thermally unstable system. While probing specific heat becomes imperative to understand local stability, information regarding global stability is divulged by the Helmholtz free energy $F$. A negative value of $F$ indicates global stability, whereas a positive value signifies global instability. The position of global phase transition occurs at $r_h^f$. It is found that an SBHD with higher $\rs$ or $r_s$ is globally stable over a wider range of $r_h$. \\
It is instructive to study optical properties such as gravitational lensing and shadow, for they bear the signature of intrinsic characteristics of the underlying spacetime. Utilizing the Gauss-Bonnet theorem, we first obtained a weak deflection angle $\gamma_D^0$ assuming a straight-line trajectory of the light ray. Then, taking the corrected trajectory with higher-order terms in M and integrating from $0$ to $\pi+\gamma_D^0$, we obtained a deflection angle with higher-order correction terms. The presence of DM is found to favorably impact the deflection angle. We then delve into gauging the impact of DM on shadow. We demonstrated DM impact in Figs. (\r{rprs}), (\r{shadow}), and in table (\r{rprsval}). The photon radius and shadow radius are found to increase with core density and core radius. \\
We then moved to our final endeavor in this article: to test the feasibility of our model and constrain DM parameters against observations related to deviation parameter $\delta$ for $M87^*$ and $Sgr A^*$. Since the shadow radius of an SBHD is always greater than a \s BH in vacuum, $\delta$ is always positive, and hence, only upper bounds on $\delta$ from EHT, Keck, and VLTI observatories are considered. Similar to the shadow radius, the presence of DM favorably impacts the deviation parameter. We have demonstrated parameter space $(\rs M^2 - r_s/M)$ for $\delta$, which clearly exhibited congruence of our model with experimental observations. These observations clearly make SBHD a viable candidate for SMBH. In our next endeavor, we would like to study quasi-periodic oscillations for SBHD, which will provide another avenue for constraining DM parameters.

\end{document}